\documentclass[journal]{IEEEtran}
\usepackage{graphicx}
\usepackage{xcolor}
\usepackage[normalem]{ulem}
\usepackage{hyperref}

\usepackage{amssymb, amsmath}

\usepackage{amsmath, bm}

\ifCLASSINFOpdf

\else

\fi

\usepackage{amsmath}

\usepackage{array}
\usepackage{fixltx2e}
\usepackage{stfloats}

\begin{document}

\title{ {Theory of Diffraction by  Holes of Arbitrary Sizes} }

\author{\IEEEauthorblockN{Mostafa Behtouei$^{1,2}$,
Luigi Faillace$^1$,
Mauro Migliorati$^{2,3}$,
Luigi Palumbo$^{2,3}$
and Bruno Spataro$^1$,
}

\IEEEauthorblockA{$^1$INFN, Laboratori Nazionali di Frascati, P.O. Box 13, I-00044 Frascati, Italy}

\IEEEauthorblockA{$^2$Dipartimento di Scienze di Base e Applicate per l'Ingegneria (SBAI), Sapienza University of Rome, Rome, Italy}

\IEEEauthorblockA{$^3$INFN/Roma1, Istituto Nazionale di Fisica Nucleare, Piazzale Aldo Moro, 2, 00185, Rome, Italy }

}

\maketitle

\IEEEtitleabstractindextext{%
\begin{abstract}
New high-gradient accelerating RF cavities are nowadays developed in several national laboratories for high-energy physics applications. Ultra high gradients, up to the order of GV/m, can be achieved by using ultra compact accelerating structures in the sub-THz regime. Nevertheless, the experimental setup for measuring the main RF parameters for such compact structures is not trivial and can easily produce errors due to lack of accuracy. Moreover, Radio-Frequency (RF) simulations for these types of cavities can require a large amount of computational time. In particular, one of the main RF parameters that needs to be evaluated and measured for the accelerating structures is the reflection coefficient. In order to obtain a fast and accurate analytical estimation, we have developed the electromagnetic theory for the calculation of the coupling of a resonant cavity with an RF waveguide. This theory is based on the Bethe's small aperture polarization approach, also developed by Collin's. In this paper, we give an exact analytical expression of the reflection coefficient as function of the physics parameters of the cavity-waveguide system, which can be applied to any geometry, material and frequency.

\end{abstract}

}

\maketitle

\IEEEdisplaynontitleabstractindextext

\IEEEpeerreviewmaketitle

\section{Introduction}

\IEEEPARstart{T}{he} concept of reflection coefficient is fundamental in high accelerating periodic structures. Electromagnetic energy may be coupled from one waveguide into another one or into a cavity resonator by a small aperture located at a suitable position in the common wall. An electromagnetic theory for the calculation of the coupling between the two devices is based on the hypothesis that the aperture is equivalent to electric or/and magnetic dipole moments \cite{ref1}. These dipole moments are proportional to the normal electric and tangential magnetic field of the incident wave, respectively. The theory was originally formulated by Bethe \cite{ref1}, then developed by Collin \cite{ref2} and modified by S. De Santis, L. Palumbo and M. Migliorati for a small hole compared to the wavelength\cite{migliorati,DeSantis}. The solution procedure was based on one of Schelkunoff's field equivalence principle \cite{ref2} in which the aperture is closed by a perfect magnetic wall. The incident field, which is chosen as the field in the absence of the aperture in the waveguide, will induce a magnetic current $J_m$ and a magnetic charge $\rho_m$ on the magnetic wall surface. These sources produce a scattered field that can be expressed as a field radiated by the dipole moment of the source distribution. The field radiated by this source into the waveguide and/or a cavity, together with the specified incident field, represents the total unique solution to the coupling problem. In our case there is a dielectric iris located between the waveguide and the cavity which is supported by the thin perfectly conducting screen. Babinet's principle \cite{ref2,jackson} allows us to relate the diffraction fields of a diffracting screen (circular wall located between the waveguide and the cavity ) to the complementary screen (aperture). The basic procedure consists in replacing the problem of diffraction conducting screen with a certain aperture as a complementary screen. The presence of the screen gives rise to transmitted and reflected fields that will be denoted as scattered field. 

We introduce a theory for diffraction by holes of arbitrary sizes. We then apply the theory to the $TM_{01}$ mode of a pill-box cavity coupled with a waveguide by a small hole but comparable to the wavelength. The amplitudes of scattered fields due to polarizability of the aperture will be determined.

\begin{figure}[h]
\centering
\includegraphics[width=2.5in]{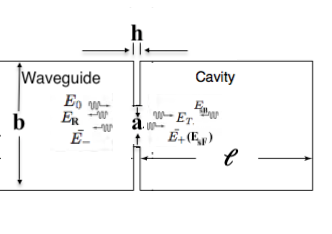}
\caption{$E_{0}$, $H_{0}$ are the primary fields in the waveguide,  $E_{sF}$, $H_{sF}$ are the scattered fields (radiation fields) due to the polarizability of the iris, and $E_{sB}$, $H_{sB}$ are the scattered fields (radiation fields) due to the polarizability of the iris from the field entered into the cavity, reflected with a phase reversal and propagated back to the iris where they become an incoming wave from the right with a certain amplitude which can polarize the iris and, due to this polarization, generate another radiation field.}
\end{figure}
Bethe's diffraction theory states that the hole is equivalent to an electric and a magnetic dipole whose moments are given by

{\small
\begin{equation}\label{1}
{\bf P_e}=\epsilon \alpha_e \hat z ({\bf E_0+E_{sF}- E_{sB}}),
\end{equation}

\begin{equation}\label{2}
{\bf P_m}=\alpha_m \hat t ({\bf H_0+H_{sF}- H_{sB}}),
\end{equation}
}
where $\alpha_e$ and $\alpha_m$ are the polarizabilites of the hole, $E_{0}$, $H_{0}$ are the primary fields in the waveguide,  $E_{sF}$, $H_{sF}$ are the scattered fields (radiation fields) due to the polarizability of the iris, and $E_{sB}$, $H_{sB}$ are the scattered fields (radiation fields) due to the polarizability of the iris from the field entered into the cavity, reflected with a phase reversal and propagated back to the iris where they become an incoming wave with a certain amplitude which can polarize the iris and, due to this polarization, generate another radiation field, as shown in Figure 1.

The Electro-Magnetic (EM) analytical method proposed in this paper can be applied for the estimation of the scattering parameters in any power-coupling network through small holes compared/comparable to the wavelength. In particular, through this approach we propose a viable tool for determining the main RF scattering parameters in particle accelerators made out of any material (metallic \cite{ref11,ref12,ref13} or dielectric \cite{ref14,ref15}, at any frequency and any temperature. For example, the reflection coefficient at the interface (such as a coupling slot) between a metallic waveguide and a resonant cavity can lose in accuracy when geometric dimensions shrink down to the order of few mm's. For example, it is not always trivial at high frequencies, especially at mm-wave range and below, to determine the scattering parameters due to the difficulty of obtaining a reliable robust measurement setup. Although these parameters can be calculated through numerical softwares, the meshing tolerances required for good accuracy can easily result in long time consuming simulations. As a specific example, we will perform the comparison between our derived analytical formulation of the reflection coefficient with the formula derived for a resonant cavity through circuit analysis \cite{ref16}.

\section{The Fields into the Structure}
The structure we are going to consider is a pillbox cavity coupled by the iris to a waveguide and the cavity will be excited in a $TM_{01}$ standing-wave mode. The starting point in the process is the incident field on the iris $a$ of Fig. (1) which polarizes the iris. The consequence of the polarization is the electric and magnetic current induction. These currents radiate scattered electric and magnetic fields into the structure (waveguide and cavity) which we have called $E_{BF}$, $H_{BF}$ and $E_{SF}$, $H_{SF}$  respectively. With the assumption that the only propagating mode inside the structure is the $TM_{01}$, the fields can be written as,

{\small
\begin{multline}\label{3}
{\bf E}={\bf E_0+E_{sF}- E_{sB}}=\xi J_0(k_c r) e^{-j \beta z}\\
+ \xi_{01}^- (\hat t\ e_{t01}  - \hat z\ e_{z01}) e^{j \beta_n z}- j \xi_{01}^+ (\hat t\ e_{t01}  +  \hat z\ e_{z01}) \ e^{-(2 \alpha \ell+j \beta z)}
\end{multline}

\begin{multline}\label{4}
{\bf H}={\bf H_0+H_{sF}- H_{sB}}= -\frac{\xi }{\eta_{01}}J'_0(k_c r)e^{-j \beta z}\\
+\xi_{01}^- (-\hat t\ h_{t01} + \hat z\ h_{z01})\ e^{j \beta_n z}   - j \xi_{01}^+ (\hat t\ h_{t01} +  \hat z\ h_{z01}) \ e^{-(2 \alpha \ell+j \beta z)}
\end{multline}
}

where,

$\beta$: Propagation constant;

$\eta_{01}$: Wave impedance of the $TM_{01}$;

$\alpha$: Attenuation constant per unit length;

$\ell$: The length of the cavity;

$e_{t01}, h_{t01} $, : Transverse modal fields for $TM_{01}$ mode;

$e_{z01}, h_{z01} $, : Longitudinal modal fields for $TM_{01}$ mode;

$\xi$: Amplitude of the incident wave;

$\xi_{01}^- $ is the $TM_{01}$ mode amplitude of scattered field propagating into the waveguide;

$\xi_{01}^+ $ is the $TM_{01}$ mode amplitude of scattered field propagating into the cavity.

The summation of incident field on the iris with the scattered fields are the new normal fields. They polarize again the aperture, and induce an electric and magnetic currents. Again the summation of the incident waves with the forward and backward scattered fields, polarize the iris aperture and induce currents, which will be the source of the new scattered fields. The strength of infinitesimal electric and magnetic polarization currents in cylindrical coordinates can be written as,
 
{\small
\begin{equation}\label{5}
{\bf P_e} =\frac{1}{r} \epsilon_0 \alpha_e \hat z E_z \delta (r-r_0)\ \delta(\theta-\theta_0)\  \delta(z-z_0)
\end{equation}
\begin{equation}\label{6}
{\bf P_m} =-\frac{1}{r} \alpha_m  \hat t H_t \delta (r-r_0)\ \delta(\theta-\theta_0)\  \delta(z-z_0)
\end{equation}
}
where,

${\bf P_e}:$ infinitesimal electric polarization currents

${\bf P_m}:$ infinitesimal magnetic polarization currents

$\epsilon_0:$ Electric permitivity

$\alpha_e:$ electric polarizability of the aperture

$\alpha_m:$ magnetic polarizability of the aperture

The electric and magnetic polarizabilities are constants that depend on the size and shape of the aperture through the fields. Substituting Eqs. (\ref{3}), $\ref{4}$ into the Eqs. ($\ref{5}$), (\ref{6}) we obtain:

{\small
\begin{multline}\label{7}
{\bf P_e}=\epsilon \alpha_e [\xi J_0(k_c r) e^{-j \beta z}-\xi_{01}^-  J_0(k_c r) e^{j \beta z} \\
-j \xi_{01}^+ J_0(k_c r)e^{-(2\alpha \ell+j \beta z)} ] \delta (r-r_0)\ \delta(\theta-\theta_0)\  \delta(z-z_0),
\end{multline}

\begin{multline}\label{8}
{\bf P_m}=\alpha_m [-\frac{\xi}{\eta_{01}}e^{-j \beta z}+\xi_{01}^- \frac{j\beta}{\eta_{01} k_c}e^{j \beta z}\\
-j \xi_{01}^+ \frac{j\beta}{\eta_{01} k_c}e^{-(2\alpha \ell+j \beta z)}J'_0(k_c r)]\ \delta (r-r_0)\ \delta(\theta-\theta_0)\  \delta(z-z_0).
\end{multline}
}
The electric and magnetic moments are related to electric and magnetic current sources, ${\bf J}$ and ${\bf M}$ as follows:

{\small
 \begin{equation}\label{9}
{\bf J}=j \omega {\bf P_e}
\end{equation}

\begin{equation}\label{10}
{\bf M}=j \omega \mu_0 {\bf P_m}
\end{equation}
}

The fields radiated from the sources propagate into the waveguide ($\bar E_-$ and $\bar H_-$)  and into the cavity ($\bar E_+$ and $\bar H_+$). They can be written as the superpositions of the modes \cite{ref5},
{\small
\begin{equation}\label{11}
{\bf E_+}= \Sigma_n \xi_n^+ {\bf E_n^+}=\Sigma_n \xi_n^+ (\hat t\ e_{tn} +  \hat z\ e_{zn})\ e^{-j \beta_n z},
\end{equation}

\begin{equation}\label{12}
{\bf H_+}= \Sigma_n \xi_n^+ {\bf H_n^+}=\Sigma_n \xi_n^+ (\hat t\ h_{tn}  +  \hat z\ h_{zn})\ e^{-j \beta_n z},
\end{equation}

\begin{equation}\label{13}
{\bf E_-}= \Sigma_n \xi_n^- {\bf E_n^-}=\Sigma_n \xi_n^- (\hat t\ e_{tn} -  \hat z\ e_{zn})\ e^{j \beta_n z},
\end{equation}

\begin{equation}\label{14}
{\bf H_-}= \Sigma_n \xi_n^- {\bf H_n^-}=\Sigma_n \xi_n^- (-\hat t\ h_{tn} +  \hat z\ h_{zn})\ e^{j \beta_n z},
\end{equation}
}
where the single index n is used to represent any possible TE or TM mode. For a given current we can determine the amplitude $\xi_n^+$ and $\xi_n^-$ by using Lorentz reciprocity theorem, which, in our case, can be written as,

{\small 
\begin{equation}\label{15}
\oint_S ( {\bf E_1} \times  {\bf H_2} \ - {\bf E_2} \times  {\bf H_1}).n ds=\int_V ({\bf J}\ .\ {\bf E_2} - {\bf M}\ .\ {\bf H_2} ) dv
\end{equation}
}
where S is a closed surface of the waveguide enclosing the volume V. $E_1, H_1$  are the scattered fields of $TM_{01}$ mode given by  Eqs. (\ref{13}) and (\ref{14}) propagating into the waveguide, $E_2, H_2$ are the the unperturbed fields inside the waveguide, $\bar J$, $\bar M$ are the electric and magnetic current sources related to strength of electric and magnetic moments (see Figure 2). 

\begin{figure}[h]
\centering
\includegraphics[width=2in]{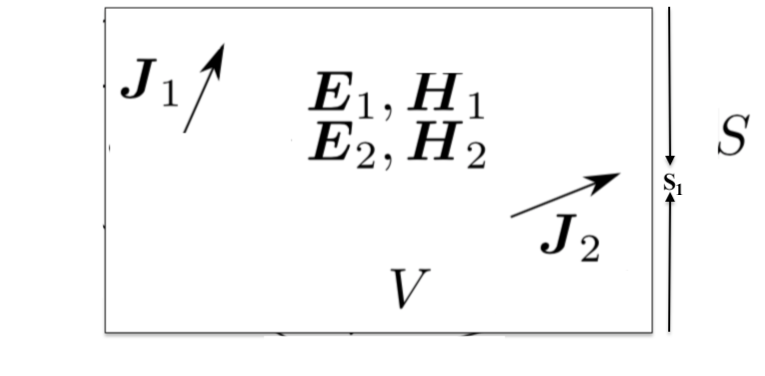}
\caption{A volume containing two sets of sources, $J_1$ and $J_2$, which each produce fields $E_1$, $H_1$ and $E_2$, $H_2$, respectively.}
\end{figure}

By applying Lorentz Reciprocity Theorem to our problem we obtain,
{\small
\begin{equation}
\oint_S ( {\bf E^-} \times {\bf H_{01}^+} \ -\ {\bf E_{01}^+} \times {\bf H^-}).\hat z ds=\int_V ({\bf J}\ .\ {\bf E_{01}^+} - {\bf M}\ .\ {\bf H_{01}^+}  )\ dv
\end{equation}
}
where ${\bf E_{01}^+}$, ${\bf H_{01}^+}$ are the fields of the $TM_{01}$ mode traveling in the positive z direction,

\begin{equation}
{\bf E_2}=  {\bf E_{01}^+}=(\hat t\ e_{t01}  + \hat z\   e_{z01})\ e^{-j \beta z},
\end{equation}

\begin{equation}
{\bf H_2}=  {\bf H_{01}^+}=(\hat t\ h_{t01}  +  \hat z\    h_{z01})\ e^{-j \beta z},
\end{equation}
where $\beta$ is the propagation constant of the $TM_{01}$ mode. With some simplifications and considering that the tangential electric field and normal magnetic field vanishes on the waveguide walls so that surface integral of the $Eq.(\ref{15})$ reduces to just integration over the cross section in which the aperture is located ($S_1$),

{\small

\begin{equation}
\xi_{01}^-\oint_{S_1}( {\bf E^-} \times {\bf H_{01}^+} \ -\ {\bf E_{01}^+} \times {\bf H^-}). \hat z ds=\int_V ({\bf J}\ .\ {\bf E_{01}^+} - {\bf M}\ .\ {\bf H_{01}^+}  ) \ dv.
\end{equation}
}
This equation can be written explicitly as,
 
 {\small
 \begin{multline}
\xi_{01}^-\int_{S_1}[(\hat t\ e_{t01}-\hat z\ e_{z01}) \times (\hat t\  h_{t01}+\hat z\ h_{z01})-(\hat t\ e_{t01}+\hat z\ e_{z01})\\
 \times (-\hat t h_{t01}+\hat z\ h_{z01})] \ .\ \hat z\   ds=\int_V ({\bf J}\ .\ {\bf E_{01}^+} - {\bf M}\ .\ {\bf H_{01}^+}  )\ dv
\end{multline}
 }
finally we obtain:

{\small
\begin{multline}
\xi_{01}^-=\frac{-1}{P_{01}} \int_V ({\bf J}\ .\ {\bf E_{01}^+} - {\bf M}\ .\ {\bf H_{01}^+}  )\ dv\\
=\frac{-1}{P_{01}} \int_V [(\hat t\ e_{t01}+\hat z\ e_{z01})\ .\ {\bf J} -(\hat t\ h_{t01}+\hat z\ h_{z01})\ .\ {\bf M}] e^{-j \beta_n z}dv
\end{multline}

where $P_{01}=2 \int_{S_1}  e_{t01} \times  h_{t01} \ .\hat z\ ds$ is a normalization constant proportional to the power flow of the $TM_{01}$ mode. Replacing ${\bf J}= j \omega {\bf P_e}$ and ${\bf M}= j \omega \mu_0 {\bf P_m}$ into the above equation we get,

 \begin{multline}
\xi_{01}^-=\frac{-1}{P_{01}} \int_V [(\hat t\ e_{t01}+\hat z\ e_{z01})\ .\ (j \omega {\bf P_e}) \\
 -(\hat t\ h_{t01}+\hat z\ h_{z01})\ .\ (j \omega \mu_0 {\bf P_m})] e^{-j \beta_n z}\ dv
\end{multline}
}
 
where $e_z(r, \phi)=(\Xi\  \sin n\phi + \xi\  \cos n\phi) J_n(k_c r)$ and $h_n(r, \phi)=\frac{-j\omega \epsilon}{k_c}(\Xi \sin n\phi+\xi \cos n\phi)J'_n(k_c r)$ are the modal fields for $TM_{010}$ mode . Replacing Eqs. $(\ref{7})$ and (\ref{8}) into the above equation we obtain,
 
\begin{multline}\label{22}
\xi_{01}^-=\frac{-1}{P_{01}} \int [(\hat t\ e_{t01}  +  \hat z\ e_{z01})\ e^{-j \beta_n z}]
  (\hat z \frac{j \omega  }{ r} \epsilon_0 \alpha_e\\  [\xi J_0(k_c r) e^{-j \beta z}-\xi_{01}^-  J_0(k_c r) e^{j \beta z}
-j \xi_{01}^+ J_0(k_c r) e^{-(2\alpha \ell+j \beta z)}])\\
 -[(\hat t\ h_{t01} + \hat z\ h_{z01})\ e^{-j \beta_n z}]\ .\ (\hat t\  \frac{j \omega  }{ r}  \alpha_m  [(-\frac{\xi}{\eta_{01}}e^{-j \beta z}\\
 +\xi_{01}^- \frac{j\beta}{\eta_{01} k_c}e^{j \beta z}-j \xi_{01}^+ \frac{j\beta}{\eta_{01} k_c}e^{-(2\alpha \ell+j \beta z)})J'_0(k_c r)]) \\
    \delta (r-r_0)\  \delta (\theta-\theta_0)\  \delta (z-z_0))  r  dr  d\theta dz.
\end{multline}

The above equation can be simplified putting the aperture at the center of the wall between the waveguide and cavity so that  we obtain

\begin{multline}
\xi_{01}^-=\frac{-1}{P_{01}} \int [(\hat t\ e_{t01}  +  \hat z\ e_{z01})\ e^{-j \beta_n z}]
  (\hat z \frac{j \omega  }{ r} \epsilon_0 \alpha_e\\  [\xi J_0(k_c r) e^{-j \beta z}-\xi_{01}^-  J_0(k_c r) e^{j \beta z}
-j \xi_{01}^+ J_0(k_c r) e^{-(2\alpha \ell+j \beta z)}])\\
 -[(\hat t\ h_{t01} + \hat z\ h_{z01})\ e^{-j \beta_n z}]\ .\ (\hat t\  \frac{j \omega  }{ r}  \alpha_m  [(-\frac{\xi}{\eta_{01}}e^{-j \beta z}\\
 +\xi_{01}^- \frac{j\beta}{\eta_{01} k_c}e^{j \beta z}-j \xi_{01}^+ \frac{j\beta}{\eta_{01} k_c}e^{-(2\alpha \ell+j \beta z)})J'_0(k_c r)]) \\
    \delta r\  \delta \theta\  \delta z)  r  dr  d\theta dz.
\end{multline}

In the next section we obtain the normalization constant proportional to the power flow of the $TM_{01}$ mode ($P_{01}$).

\section{The power flow of the nth mode}

As we have mentioned earlier $P_{01}$ is the normalization constant proportional to the power flow of the $TM_{01}$ mode and it's equal to $2 \int_{S_1} \bar e_{t01} \times \bar h_{t01}^* \ .\hat z\ ds$. TM modal fields as we have shown in Table 1, can be written as,

\begin{equation}
e_n(r, \phi)=\frac{-j\beta}{k_c}(\Xi\ \sin n\phi + \xi\ \cos n\phi) J'_n(k_c r)
\end{equation}

\begin{multline}
h_n(r, \phi)=\frac{-j\omega \epsilon}{k_c}(\Xi\ \sin n\phi+\xi\ \cos n\phi)J'_n(k_c r)\\
=\frac{-j\beta}{\eta_{01} k_c}(\Xi\ \sin n\phi+\xi\ \cos n\phi)J'_n(k_c r)
\end{multline}
where $k_c$ is cutoff wavenumber, n is the number of zeros in $E_z$ along the radial direction, $J'_n(k_c r)$ is the derivative of the Bessel function of the first kind and $\eta_{01}$ has an exactly defined value $\eta_{01}=376.73$. For $TM_{01}$ modal fields we can write,

\begin{equation}
e_{t01}=\frac{-j\ \beta}{k_c} J'_0(k_c r)
\end{equation}

\begin{equation}
h_{t01}=\frac{-j\ \beta}{\eta_{01} k_c}  J'_0(k_c r)
\end{equation}

Substituting the two equations above into  $P_{01}$ we have,
{\small
\begin{multline}
P_{01}=2 \int_{z}  e_{01} \times  h_{01} \ .\hat z ds\\
=-\frac{2\beta^2}{\eta_{01} k_c^2 }\ \int_0^{2\pi}\ \int_0^b\ J'^2_0(k_c r)\ r\ dr\ d\theta
=-\frac{4 \pi  \beta^2}{\eta_{01} k_c^2}\  \int_0^b\ r J'^2_0(k_c r)\ dr
\end{multline}

}
 where we evaluated the integration over the cross section  of the cavity. Since $J'_0(k_c r)=-J_1(k_c r)$ and replacing $k_c r=\chi_{01}$ we obtain,
 
 {\small
 \begin{equation}\label{29}
P_{01}=\frac{2 \pi  b^2 \beta^2 }{\eta_{01} k_c^2}\ J_1^2(\chi_{01})=\frac{2 \pi  b^2  }{\eta_{01} } [1-(\frac{k}{k_c})^2] J_1^2(\chi_{01})
\end{equation}

}

\begin{table}
\centering
 \begin{tabular}{||c| c||} 
 \hline
 Quantity &  $TM_{nm}$ Mode \\ [0.5ex] 
 \hline\hline
k & $\omega \sqrt{\mu \epsilon}$   \\
 \hline
$k_c$&   $\frac{Xi'_{nm}}{b}$   \\ 
 \hline
$\beta$ &  $\sqrt{k^2-k_c^2}$  \\
 \hline
$\lambda_c$&  $\frac{2\pi}{k_c}$  \\
  \hline
$\lambda_g$&  $\frac{2\pi}{\beta}$  \\
 \hline
 $\nu_p$& $\frac{\omega}{\beta}$   \\
 \hline
  $\nu_g$&  $\frac{d \omega}{d \beta}$   \\
 \hline
$\alpha_d$& $\frac{k^2 tan \delta}{2 \beta}$  \\
 \hline
$E_z$&  $(\Xi sin n\phi + \xi cos n \phi)\ J_n(k_c r)\ e^{-j \beta z}$  \\ 
 \hline
 $H_z$&  0  \\[1ex] 
 \hline
 $E_r$& $\frac{-j \beta}{k_c}\ (\Xi sin n\phi + \xi cos n \phi)\ J'_n(k_c r)\ e^{-j \beta z}$  \\ 
 \hline
 $E_\phi$&  $\frac{-j \beta n}{k_c^2 r}\ (\Xi sin n\phi - \xi cos n \phi)\ J'_n(k_c r)\ e^{-j \beta z}$  \\[1ex] 
 \hline
 $H_r$& $\frac{-j \omega \epsilon n}{k_c^2 r}\ (\Xi sin n\phi - \xi cos n \phi)\ J'_n(k_c r)\ e^{-j \beta z}$   \\ 
 \hline
 $H_\phi$&  $\frac{-j \omega \epsilon }{k_c}\ (\Xi sin n\phi + \xi cos n \phi)\ J'_n(k_c r)\ e^{-j \beta z}$   \\[1ex] 
 \hline
$\eta$ &  $\eta_{TM}=\frac{ \beta \eta}{k}$  \\[1ex] 
 \hline
\end{tabular}
\caption{Summary of Results for the wave propagation of the $TM_{nm}$ mode \cite{ref4}.}
\end{table}

\section{Calculation of electric polarizability}

According to Bethe's theory, the electric polarization coefficient is $\alpha=-2/3 a^3$ \cite{ref1}. The assumption for using this coefficient is that the holes should be small compared with the wavelength. Based on this theory for the bigger holes - the size is comparable to the wavelength - we should consider the variation of the normal electric field in the Green's function which is in a factor of $e^{i k a}$ and this correction can be of the order of $(ka)^2$ rather than $ka$. Eggimann \cite{ref5} solved the problem of diffraction of arbitrary electromagnetic field by a circular perfectly conducting disk using a series representation in powers of k using the results of generalized Babinet's principle\cite{ref2} by considering equivalence of the disk and the aperture problems we have

{\small
\begin{equation}
 P_z=\frac{4}{3} a^3 \epsilon_0 (1- \frac{3}{10}(ka)^2 + j \frac{4}{9\pi}   (ka)^3) E_z^0+\frac{2a^5 \epsilon_0}{15}\ \frac{\partial^2 E_z^0}{\partial z^2}
 \end{equation}
 
 \begin{equation}
=\frac{4}{3} a^3 \epsilon_0  E_z^0- \frac{2}{5} a^3 \epsilon_0 (ka)^2 E_z^0+\frac{2a^5 \epsilon_0}{15}\ \frac{\partial^2 E_z^0}{\partial z^2}+j \frac{16}{27\pi} a^3 \epsilon_0 (ka)^3 E_z^0.
 \end{equation}

 }
 
For the small holes compared to the wavelength the second, third and fourth terms vanish as the radius of the holes is in the order of 6, 5 and 6 respectively. One can derive the electric polarizability of Bethe's theory by small holes.  For the bigger holes all terms have contribution on electric polarizability. It should be noted that the imaginary part of the equation means that electric moment is out of phase with the electric field and there is a phase delay delay between the application of the electric field and the induced dipole moment. The imaginary term is responsible for $\pi/2$ phase shift of the iris located  between cavity resonators. The third term contribution where we have the second derivative of the electric field is field oscillations and converting this term in time domain we get a term with a $\frac{1}{c^2}$ coefficient which destroys the effect of the second derivatives and the equation becomes

{\small
  \begin{equation}
 P_z=\frac{4}{3} a^3 \epsilon_0 (1- \frac{3}{10}(ka)^2 + j \frac{4}{9\pi}   (ka)^3) E_z^0.
 \end{equation}
 
 }

Where the electric polarization coefficient can be obtained

{\small
  \begin{equation}\label{34}
\alpha=\frac{4}{3} a^3 \epsilon_0 (1- \frac{3}{10}(ka)^2 + j \frac{4}{9\pi}   (ka)^3).
 \end{equation}
 
 }
We observe a term involving $(ka)^2$ for the hole size comparable to the wavelength. This is exactly the correction mentioned in Bethe's theory for the diffraction by bigger holes.  

For the small irises all the modes of waveguide are below cut-off and their attenuation is exponential with respect to the length of the iris. McDonald\cite{ref6} has derived an electric polarizability in which the attenuation term can be considered and it is as follows,

{\scriptsize
\begin{equation}\label{31}
 \alpha_e=C_E\ \alpha_e^*\ e^{-2.405h/a} \ \ \ with \ \ C_E=0.83\ \ and\ \ \alpha_e^*=-\frac{2}{3} a^3
\end{equation}

}
where h is the iris thickness, $C_E$ is the numerical value calculated by McDonald. 

Gluckstern \cite{ref7} has derived integral equations for the potential and field distribution within a circular hole in a plane conducting wall of finite thickness induced by uniform field. Then he obtained variational expressions for the polarizability and susceptibility of the hole, from which one can obtain the electric and magnetic dipole moments induced on the inside (far field) and outside (no far field) boundaries of the hole \cite{ref7}. His results had a good agreement with McDonald's results.

One can use the Eq. (\ref{34}) when the holes are comparable to the wavelength. In the next section we will calculate the amplitudes of scattered fields due to polarizability of the aperture located between waveguide and the cavity will find equations for reflection and transmission coefficients.

\section{Reflection  Coefficient}

Finally we can calculate the amplitude for forward and backward waves due to dipole moments,

 \begin{multline}\label{33}
\xi_{01}^-=\frac{-1}{P_{01}}  j  \omega  \epsilon_0 \alpha_e \ \xi+ \frac{-1}{P_{01}} \int_0^a \int_0^{2\pi} \int_0^h j \omega  \epsilon_0 \alpha_e  \xi_{01}^- \\
( - J_0^2(k_c r))e^{j \beta z} \delta r  \delta \theta \delta z  dr \ d\theta \ dz\\
+ \frac{-1}{P_{01}} \int_0^a \int_0^{2\pi} \int_0^h j \omega  \epsilon_0 \alpha_e\\
  [-j \xi_{01}^+ ( J_0^2(k_c r))e^{-2\alpha \ell-j \beta z}   ])\\
   \delta r  \delta \theta \delta z  dr \ d\theta \ dz,
\end{multline}

where the electric polarization coefficient is given

{\small
  \begin{equation}
\alpha_e=\frac{4}{3} a^3 \epsilon_0 (1- \frac{3}{10}(ka)^2 + j \frac{4}{9\pi}   (ka)^3)
 \end{equation}
  }
and  $P_{01}=\frac{2 \pi b^2  }{\eta_{01} } [1-(\frac{k}{k_c})^2] J_1^2(\chi_{01})$ is the normalization constant proportional to the power flow of the $TM_{01}$ mode.  The imaginary part of the electric polarizability  $\alpha_e$ means that electric moment is out of phase with the electric field and there is a phase delay delay between the application of the electric field and the induced dipole moment. The imaginary term is responsible for $\pi/2$ phase shift of the iris located  between cavity resonators. Finally $\xi_{01}^-$ and $\xi_{01}^+$  are the amplitudes of the radiation fields due to iris as a dipole. They can be written as reflection and transmission coefficients.

\begin{equation}
\xi_{01}^-=\Gamma\  \xi,
\end{equation}

\begin{equation}
\xi_{01}^+=(1-\Gamma)\  \xi.
\end{equation}

Replacing these equations into the equation above we get,

\begin{multline}
\Gamma =\frac{-1}{P_{01}}  j  \omega  \epsilon_0 \alpha_e  (1-j\ e^{-2\alpha \ell} )+ \frac{-1}{P_{01}} \int_0^a \int_0^{2\pi} \int_0^h j \omega  \epsilon_0 \alpha_e  \Gamma\\
 (- J_0^2(k_c r)) e^{j \beta z}(1+j\ e^{-2\alpha \ell-2j \beta z} )\\
  \ \delta r\   \delta \theta\  \delta z\   dr \ d\theta \ dz,
\end{multline}

writing the equation above in a compact way we obtain,

{\small
\begin{equation}\label{37}
{\Gamma = \Gamma_0+ D\int_V \Gamma\  f(r) g(z) h(\theta)\delta r\ \delta \theta\  \delta z\   dV},
\end{equation}
}
where 

$\Gamma_0=\frac{-1}{P_{01}}  j  \omega  \epsilon_0 \alpha_e  (1-j\ e^{-2\alpha \ell} )$,

$D=\frac{-1}{P_{01} }  j  \omega  \epsilon_0 \alpha_e $,

$f(r)= J_0^2(k_c r)$,

$g(z)=e^{j \beta z}(1+j\ e^{-2\alpha \ell-2j \beta z} ) $,

and 

$h(\theta)$: Heaviside step function.

To obtain equation above we assumed that the iris is at the center of the cross section of the cavity. This assumption simplify the equation of infinitesimal electric and magnetic moments (Eq.s ($\ref{5}$) and ($\ref{6}$) to the equations below,

\begin{equation}
\bar{P_e} =\frac{1}{r} \epsilon_0 \alpha_e \hat n E_n \delta (r)\ \delta(\theta)\  \delta(z),
\end{equation}

\begin{equation}
\bar{P_m} =-\frac{1}{r} \alpha_m \bar{H_t} \delta (r)\ \delta(\theta)\  \delta(z),
\end{equation}

these equations allow us make a convolution of the functions at the center and simplify the equations. By considering the aperture at the center of the wall between the waveguide and the cavity and using self consistent solution method we obtain:

\begin{equation}
\Gamma= \frac{\frac{-1}{P_{01}}  j  \omega  \epsilon_0 \alpha_e  (1-j\ e^{-2\alpha \ell} )}{1+j\frac{-1}{P_{01} }   \omega  \epsilon_0 \alpha_e (1+j\ e^{-2\alpha \ell}) },
\end{equation}

where $P_{01}$ is a normalization constant proportional to the power flow of the $TM_{010}$ mode

\begin{equation}
P_{01}= \frac{2 \pi b^2  }{\eta_{01} } [1-(\frac{k}{k_c})^2] J_1^2(\chi_{01}),
\end{equation}

and  $\alpha_e$ is the electric polarizability 

{\small
  \begin{equation}
\alpha_e=\frac{4}{3} a^3 \epsilon_0 (1- \frac{3}{10}(ka)^2 + j \frac{4}{9\pi}   (ka)^3).
 \end{equation}
  }
The imaginary part of the electric polarizability  $\alpha_e$ means that electric moment is out of phase with the electric field and there is a phase delay delay between the application of the electric field and the induced dipole moment. Finally $\xi_{01}^-$ and $\xi_{01}^+$  are the amplitudes of the radiation fields due to iris as a dipole. This equation is an analogous of the reflection coefficient obtained by Collin for $TE_{10}$ using the circuit theory which for a comparison we report here \cite{ref2},

\begin{equation}
1-\frac{4j \alpha_m \beta_{10}/ab  }{1+jX+W},
\end{equation}

Where $X=2 \alpha_m \beta_{10}/ab$ and

\begin{equation}
W=-\frac{4 \alpha_m k_0^2 \pi^2}{k_{101}^2 abd^3[k_{101}^2-k_0(1+\frac{1-j}{Q})]}
\end{equation}

To simplify Eq. ($\ref{22}$) we observed that the magnetic moment has no contribution for the $TM_{01}$ mode and the second term in the equation can be totally eliminated. In the following we will prove that considering the magnetic moment for $TM_{01}$ mode leads to the zero contribution for reflection coefficient calculation. Eq. (\ref{22}) for the magnetic moment part becomes

 \begin{multline}
\xi_n^-=\frac{1}{P_{01}} \int_V (\bar h_{t01}+\hat z h_{z01})\ .\ (j \omega \mu_0 \bar P_m) e^{-j \beta_n z}dv
\end{multline}

substituting Eq. (\ref{8}) into the above equation we obtain

\begin{multline}
\xi_{01}^-=\frac{1}{P_{01}} \int [(\bar{h_{t01} }+\  \hat z h_{z01}) e^{-j \beta_n z}]\\
 (\hat t\  \frac{j \omega  }{ r}  \alpha_m  (-\frac{\xi}{\eta_{01}}e^{-j \beta z} +\xi_{01}^- \frac{j\beta}{\eta_{01} k_c}e^{j \beta z}\\
  -j \xi_{01}^+ \frac{j\beta}{\eta_{01} k_c}e^{-(2\alpha \ell+j \beta z)})J'_0(k_c r))   \delta r\  \delta \theta\  \delta z r  dr  d\theta dz.
\end{multline}

Due to the properties of the Dirac delta function, the derivative of the Bessel function of the zero kind and first kind at the center of the holes is zero and leads to  $\xi_{01}^-$=0, as a consequence, for the $TM_{01}$ mode, the iris aperture between the pill-box cavity coupled with a cylindrical waveguide, acts as an electric moment.

\section{Conclusion}

We introduced a theory for diffraction by holes of arbitrary sizes. We then applied this theory to the $TM_{01}$ mode cavity coupled with a cylindrical waveguide by a small hole comparable to the wavelength. The amplitudes of scattered fields due to polarizability of the aperture located between waveguide and the cavity have been determined and we have found equations for reflection and transmission coefficients. The Electro-Magnetic (EM) analytical method proposed in this paper can be applied for the estimation of the scattering parameters in any power-coupling network through small holes.

\end{document}